\documentclass[titlepage,twoside,12pt]{article}
\usepackage{amssymb}
\usepackage{amsfonts}
\begin{document}

{\bf \Large Where and How Do They Happen ?} \\ \\

{\bf Elem\'{e}r E Rosinger} \\ \\
Department of Mathematics \\
and Applied Mathematics \\
University of Pretoria \\
Pretoria \\
0002 South Africa \\
eerosinger@hotmail.com \\ \\

{\bf Abstract} \\

This is a two part paper. The first part, written somewhat earlier, presented standard processes
which cannot so easily be accommodated within what are presently considered as physical type
realms. The second part further elaborates on that fact. In particular, it is argued that quantum
superposition and entanglement may better be understood in extensions of what we usually consider
as physical type realms, realms which, as it happens, have so far never been defined precisely
enough. \\ \\

{\bf \Large Part I} \\ \\

{\bf Abstract} \\

It has for ages been a rather constant feature of thinking in science to take it for granted that
the respective thinking happens in realms which are totally outside and independent of all the
other phenomena that constitute the objects of such thinking. The imposition of this divide on
two levels may conflict with basic assumptions of Newtonian and Einsteinian mechanics, as well as
with those in Quantum Mechanics. It also raises the question whether the realms in which thinking
happens have no any other connection with the realms science deals with, except to host and allow
scientific thinking. \\ \\

{\bf 0. The Yet Undefined Physical Realms ...} \\

In the sequel, based on rather obvious and simple, even if so far seldom considered facts
within, or related to Physics, we shall argue that what are usually assumed to be the Physical
Realms may have to be extended. Such possible additional realms, however, are not along those
infinitely many of Everett's "many-worlds" view of Quantum Mechanics. Instead, they are
suggesting a finite number of further physical type realms, thus they can be seen as a
development of the classical Cartesian realm of "res extensa". \\

As for what Physical Realms may actually mean, or rather, Physics itself, here is a recent and
quite appropriate view on that never yet clarified issue, [8, pp. 153,154] :

\begin{quote}

"Physics is the study of those phenomena that are successfully treatable with well-specified
and testable models. \\
For example, Physics treats atoms and simple molecules. Chemistry, on the other hand, deals
with all molecules, most of whose electron distributions cannot be well specified. A
physicist might study a well specified biological system, but the functioning of a complex
organism lies in the domain of biologists. \\
Anything not successfully treatable with a well-specified and testable model is rather quickly
defined out of Physics."

\end{quote}

It is quite clear in this spirit that, even if no one seems to care much about a more precise
definition of Physics, and thus, of Physical Realms, phenomena such as human thinking, let
alone, human consciousness or awareness, are not expected to concern Physics any time soon.
Consequently, what for Descartes constituted "res cogitans", that is, the realms of thinking,
are supposed to remain in the splendour of their undisturbed solitude, as far as Physics is
concerned. And then, anything that may be seen as remotely acceptable from a physical point of
view, may be but a refinement, or rather, a structural enrichment of the Cartesian "res
extensa", that is, of the realms which, at least intuitively, are supposed to have to do with
Physics. \\

And yet, as seen in the sequel, the story is not quite that simple, not even from a strictly
physical point of view ... \\ \\

{\bf 1. Conflict with Newtonian Mechanics} \\

Instant action at arbitrary distance, such as in the case of gravitation, is one of the basic
assumptions of Newtonian mechanics. This certainly does not appear to conflict with the fact that
we can think instantly and simultaneously about phenomena which are no matter how far apart from
one another in space or in time. However, absolute space is also a basic assumption of Newtonian
Mechanics. And it is supposed to contain absolutely everything that may exist in Creation, be it
in the past, present or future. Consequently, it is supposed to contain, among others, the
physical body of the thinking scientist as well. \\

Yet it is not equally clear whether it also contains scientific thinking itself which,
traditionally, is assumed to be totally outside and independent of all phenomena under its
consideration, therefore in particular, of the Newtonian absolute space, and also, of absolute
time. \\

And then the question arises :

\begin{quote}

Where and how does such a scientific thinking take place or happen ?

\end{quote}

{~}\\

{\bf 2. A difference with Mathematics} \\

Mathematical thinking, especially in its modern and abstract variants, does not appear to need
the assumption of any absolute space, or for that matter, absolute time. Such thinking may appear
to unfold during appropriate local time intervals. However, when seen all in itself, and
unrelated to the physical body of the respective mathematician, it is quite likely that such
thinking has no location in any space, be it relative or absolute. \\ \\

{\bf 3. Conflict with Einsteinian Mechanics} \\

In Einsteinian Mechanics a basic assumption is that there cannot be any propagation of action
faster than light. \\
Yet just like in the case we happen to think in terms of Newtonian Mechanics, our thinking in
terms of Einsteinian Mechanics can again instantly and simultaneously be about phenomena no
matter how far apart from one another in space or time. \\

Consequently, the question arises :

\begin{quote}

Given the mentioned relativistic limitation, how and where does such a thinking happen ?

\end{quote}

{\bf 4. Conflict with Quantum Mechanics} \\

Let us consider the classical EPR, or Einstein-Podolsky-Rosen entanglement phenomenon, and for
simplicity, do so in the terms of quantum computation. For that purpose it suffices to
consider double qubits, that is, elements of $\mathbb{C}^2 \bigotimes \mathbb{C}^2$, such as
for instance
the EPR pair \\

(4.1)~~~ $ \begin{array}{l}
               |~ \omega_{00} > ~=~ |~ 0, 0 > ~+~ |~ 1, 1 > ~=~ \\ \\
                ~~~=~ |~ 0 > \bigotimes |~ 0 > ~+~ |~ 1 > \bigotimes |~ 1 >
                         \, \in  \mathbb{C}^2 \bigotimes \mathbb{C}^2
            \end{array} $ \\

which is well known to be {\it entangled}, in other words, $|~ \omega_{00} > $ is {\it not} of
the form \\

$~~~~~~ ( \alpha |~ 0 > ~+~ \beta |~ 1 > ) \bigotimes ( \gamma |~ 0 > ~+~ \delta |~ 1 > )
                                           \, \in  \mathbb{C}^2 \bigotimes \mathbb{C}^2 $ \\

for any $\alpha, \beta, \gamma,\delta \in \mathbb{C}^2$. \\

Here we can turn to the usual and rather picturesque description used in quantum computation,
where two fictitious personages, Alice and Bob, are supposed to exchange information, be it of
classical or quantum type. \\
Alice and Bob can each take their respective qubit from the entangled, or EPR pair of qubits
$|~ \omega_{00} >$, and then go away with it no matter how far apart from one another. And the
two qubits thus separated in space will remain entangled, unless of course one or both of them
get involved in further classical or quantum interactions. For clarity, however, we should
note that the single qubits which, respectively, Alice and Bob take away with them from the
EPR pair $|~ \omega_{00} >$ are neither one of the terms $|~ 0, 0 >$ or $|~ 1, 1 >$ in (4.1),
since both these are themselves already pairs of qubits, thus they cannot be taken away as
mere single qubits, either by Alice, or by Bob. Consequently, the single qubits which Alice
and Bob take away with them cannot be described in any other form, except that which is
implicit in (4.1). \\

Now, after that short detour into the language of quantum computation, we can note that,
according to Quantum Mechanics, the entanglement in the EPR double qubit $|~ \omega_{00} >$
implies that the states of the two qubits which compose it are correlated, no matter how far
from one another Alice and Bob would be with them. Consequently, knowing the state of one of
these two qubits can give information about the state of the other qubit. On the other hand,
in view of General, or even Special Relativity, such a knowledge, say by Alice, cannot be
communicated to Bob faster than the velocity of light. \\

And yet, anybody who is familiar enough with Quantum Mechanics, can instantly know and
understand all of the above, no matter how far away from one another Alice and Bob may be with
their respective single but entangled qubits. \\

So that, again, the question arises :

\begin{quote}

How and where does such a thinking happen ?

\end{quote}

{~} \\

{\bf 5. Two, Among Other Possible Alternatives} \\

Let us first assume that scientific thinking does indeed happen in realms outside and
independent of all the realms in which the variety of phenomena studied by scientific thinking
takes place. Then the very existence of scientific thinking proves the existence of realms
transcendental to those which at present are customarily the object of that scientific
thinking. \\
In this case, one may ask whether the realms in which scientific thinking happens have, indeed,
no any other connection whatsoever with the realms which are the object of study of science,
except to host and allow such scientific thinking. \\

A cautious answer is of course not one of categorical negation. Furthermore, any answer,
including a categorically negative one, may need some supporting evidence, and possibly of
experimental or empirical kind as well. \\

If alternatively, we assume that, after all, there is only one overall realm in which
everything happens, then quite likely we may have to extend rather significantly, if not in
fact dramatically, the list of entities, phenomena, or processes which are, or can be relevant
in Physics, Chemistry, Biology, and so on. Certainly, in such a case it can no longer be taken
for granted - and done so without any supporting evidence - that the whole range of entities
and their interactions which form the object of science are isolated in some subdomain of that
unique overall realm. And very much isolated they appear to be, since usual scientific
thinking itself is assumed to be outside and independent of them, plus we deal with all those
entities and their interactions as if they were perfectly self-contained. \\ \\

{\bf 6. Conclusions} \\

It may be useful to ask the following four questions :

\begin{quote}

1. Do we believe that whatever in Creation which may be relevant to science is already
accessible to our awareness ? \\

2. And if not - which is most likely the case - then do we believe that it may become
accessible during the lifetime of our own generation ? \\

3. And if not - which again is most likely the case - then do we believe that we should
nevertheless try some sort of two way interactions with all that which may never ever become
accessible to the awareness of our generation, yet may nevertheless be relevant to science
even in our own days ? \\

4. And if yes - which most likely is the minimally wise approach - then how do we intend to
get into a two way interaction with all those realms about which our only awareness can be
that they shall never ever be within our awareness, or perhaps, not even of human awareness as
such, no matter how long our species may live ?

\end{quote}

{~} \\

{\bf \Large Part II} \\ \\

{\bf Abstract} \\

It is further argued that quantum superposition and entanglement may better be understood in
extensions of what we usually consider to be physical type realms, realms which in fact have
never been defined precisely enough. \\ \\

{\bf 1. Superposition : a Typically Quantum \\
\hspace*{0.4cm} Fundamental Phenomenon} \\

In the non-relativistic Quantum Mechanics of a finite system $S$ described by states in a
Hilbert space $H$, if for instance $\psi_1,~ \psi_2 \in H$ are two possible orthogonal states
of the system $S$, then further states of $S$ are given by the arbitrary linear
combinations \\

(1.1)~~~ $ \psi = c_1 \psi_1 + c_2 \psi_2,~~~ c_1, c_2 \in \mathbb{C} $ \\

where the usual normalizing conditions are assumed \\

(1.2)~~~ $ || \psi_1 || = || \psi_2 || = 1,~~~ | c_1 |^2 + | c_2 |^2 = 1 $ \\

hence resulting as well in \\

(1.3)~~~ $ || \psi || = 1 $ \\

So far, in no other theory of Physics is such a property present. As for the importance of
that property in Quantum Mechanics it suffices to recall two facts : it leads to yet unsolved
foundational controversies, as in the celebrated argument in Schr\"{o}dinger's Cat, and
it is considered to be one of the basic resources of quantum computers, a resource which
allows them unprecedented computational power, a power not possible to attain with usual
electronic computers. \\ \\

{\bf 2. Realms Physical, and Other Ones Less So ?} \\

It is nowadays a fundamental assumption that the Physical Realms do surely contain all there
is, or at least, all there is of interest to Physics. And as with many a fundamental
assumption, this one is so deeply ingrained that hardly anyone finds any reason at all to make
it explicit to any extent. \\
One of the amusing aspects of such an approach is the convenient {\it circularity} of the
argument, a circularity which, however, does not seem to concern in the least its
proponents ... \\
Another amusing aspect is the recently emerging credo, according to which "information is
physical" ... \\
This credo does, of course, reflect an awareness that what earlier were perceived, mostly
tacitly, as the possible boundaries of the Physical Realms should now be extended in order not
to leave out such an entity of fast growing importance like information. \\
And needless to say, such a move to encompass information within the Physical Realms is rather
easy to accomplish, since the latter remains as undefined as it has always been ... \\
Indeed, there is here a significant mismatch between the rather clear definition of
information in present day science and technology, and on the other hand, the actual, and
quite convenient vagueness of what the Physical Realms are supposed to be about. Not to
mention that, in spite of the insistent propagation of that newly emerged credo, the concept
of information is in fact treated as a second class one at best in most of the present day
fundamental theories of Physics, including in Quantum Mechanics. \\

On the other hand, and despite of the above, it is quite clear that the so called Physical
Realms, even in their ever vague and latest extended sense, do {\it not} contain all that is
of interest. And on top of it, they happen to {\it fail} to do so precisely on their own
terms. \\
Several such instances were discussed in Part I, and here we recall one of them :

\begin{quote}

Anybody, and even more so a physicist, can at the same time think about two arbitrarily far
away places in the universe, for instance, two galaxies. \\
On the other hand, according to Relativity Theory, no physical interaction can take place
with arbitrary velocity. \\
Thus such a thinking, so easily and so commonly available to quite everybody, {\it cannot} be
of a physical nature.

\end{quote}

And then, the question arises :

\begin{quote}

Where and how does such a thinking happen, if not within the Physical Realms, and definitely
not there, in view of Physics itself ?

\end{quote}

And while such a question remains unanswered, and in fact, not even considered by present day
Physics, perhaps one may as well compound the issue with the following. \\

It was Descartes in the early 1600s, who suggested the existence of two distinct realms. His
"res extensa" was more or less what has been meant by the Physical Realms. On the other hand,
his "res cogitans" was a realm beyond and outside of "res extensa", and it encompassed
thinking. \\
As a consequence, Descartes has for long been ridiculed as being a dualist ... \\
Such a judgment misses, however, the fundamental fact that, as so many major European
scientists of his time, Descartes himself was a deeply religious person in the Christian
tradition. Consequently, he could not possibly be less removed from dualism than anybody else,
since he saw God as underlying all Creation, and thus in particular, both "res extensa" and
"res cogitans". \\

Now of course, Descartes himself did not advocate the study of "res cogitans" by the means of
Physics, whatever the latter may mean under reasonable conditions. \\
And Classical Physics, that is, prior to the 20th century, did not in any way seem to require
a more direct involvement of "res cogitans" than it would usually happen in the customary
thinking process of normal humans, among them, physicists. \\

Relativity Theory, in spite of the above question, has not changed that classical situation,
and it did not appear to need to do so. What it does instead, and even if not yet seriously
considered, is to point quite sharply to the existence of at least two very different realms.
And for the lack of better terms, as well as a homage to Descartes, we can still call those
two realms as "res extensa" and "res cogitans", respectively. \\ \\

{\bf 3. Does Superposition Need a Third Realm ?} \\ \\

This may not be such an easy to answer question as one would like it. Indeed,
Schr\"{o}dinger's Cat already shows that it is not trivial. Therefore, let us consider it with
some care. \\

What is obvious from (1.1) - (1.3) is that superposition takes place in the Hilbert space $H$,
that is, within the mathematical model of the quantum system $S$. And as mathematical models
go, they may hopefully reflect their respective system which, of course, is supposed to be
situated in "res extensa", but on the other hand, as mere models, are {\it not} supposed to be
identical with such a system. \\
This failure to distinguish between a physical system and its model is precisely one of the
reasons one ends up with the controversy about Schr\"{o}dinger's Cat. \\

And then, the question arises :

\begin{quote}

Are superpositions (1.1) - (1.3) bona fide physical phenomena, or on the contrary, they are
merely convenient features of the respective mathematical model ?

\end{quote}

Well, as far as one can understand, this question does not have a clear enough answer in
present day Quantum Mechanics. \\

However, as it may happen with not a few physicists, in case one tends to consider
superpositions as genuine physical phenomena, then the foundational controversy around
Schr\"{o}dinger's Cat may simply be set aside by considering a {\it third} realm which we may
call "res super-extensa", and in which such superpositions take place. This realm contains the
usual "res extensa", in the sense that $\psi$ in (1.1), as a superposition of  $\psi_1$ and
$\psi_2$, belongs to it, without however belonging to "res extensa", while $\psi_1$ and
$\psi_2$ belong to the latter. Clearly, just as with "res extensa", there is no need for any
overlapping between "res super-extensa" and "res cogitans". \\

Here, however, one should note that the mathematical model (1.1) - (1.3), assumed to be in
"res cogitans", need not always distinguish between "res extensa" and "res super-extensa".
Indeed, $\psi$ in (1.1), as an element of the Hilbert space $H$, can in itself belong to "res
extensa", as long as it is not seen as being constituted as a superposition. \\

The point to note with the above is that it is precisely the preference to see superpositions
as physically real, that is, as having genuine physical existence, and not merely being
representations in a mathematical model, which, when considered together with conundrums such
as Schr\"{o}dinger's Cat, can suggest the consideration of a third realm, such as that of "res
super-extensa". \\ \\

{\bf 4. And How About Entanglements ?} \\

As seen in Part I, entanglement also raises a question as to where and how it happens, given
what appears to be its instantaneous nonlocal manifestation. \\

And yet, it may appear that entanglement, even more than superposition, is seen by physicists
as a genuine physical phenomenon, and not merely as some occurrence in the mathematical
model. \\
In this regard, no less than superpositions, entanglements are typical quantum phenomena, as
well as unprecedented resources in quantum computation. As for their foundational importance,
it suffices to recall the celebrated EPR paper, with all the related subsequent
developments. \\

Thus a fundamental and still controversial issue which entanglements bring up is that of {\it
nonlocality}. This fact, as is well known, was brought forward most starkly with the
celebrated Bell Inequalities. \\

Here however, once one may consider the possibility of a third realm, like for instance, the
above "res super-extensa", which is in fact but a {\it larger} instance of the customary "res
extensa", the very issue of nonlocality may benefit from a new view and understanding. \\
Indeed, it may simply happen that in "res super-extensa" the dichotomy "local - nonlocal" is
meaningless. \\
And here we should recall that such a possibility is not at all strange, since in a bounded
system modelled mathematically by a compact space, the very concept of "nonlocal" loses much
of its usual difficulties, if not in fact, its meaning. And in this regard we can recall that,
so far, the very question whether the whole of the universe itself is in fact bounded is still
open. \\

But then, and as if to complicate the issues, entanglements need {\it not} necessarily happen
in the same extension of "res extensa" in which superposition may happen. Consequently, we may
yet have to consider another, namely, third physical type realm as well. \\

A further possible consequence of considering physical extensions of the usual "res extensa"
is that the foundational controversy related to the so called "hidden variables" in Quantum
Mechanics may give way in favour of whole physical realms which, so far, were themselves
hidden. In other words, it may well happen that what has been missing were not some hidden
variables within this or that quantum entity, but rather whole physical type {\it realms}
within which the very quantum processes as a whole may actually take place. \\

And with the acceptance in String Theory of the fact that the so called Physical Realms may
have highly counterintuitive large finite dimensions, some of them so contracted as to make
the dichotomy "local - nonlocal" quite meaningless, there is no longer any particular reason
to be so parsimonious when considering the possible realms, beyond the usual "res extensa",
that may be relevant to Physics. \\ \\

{\bf 5. Conclusions} \\

Several extensions of what usually is meant by the otherwise undefined concept of Physical
Realms were argued, based on rather obvious, simple, as well as fundamental physical
considerations. In section 2, such an extension is motivated by the limitation of velocity of
physical interactions, as follows from Relativity. In section 3, it was argued that, precisely
to the extent that quantum superposition is not a mere feature of a mathematical model, but a
genuine physical phenomenon, an extension of the customary concept of Physical Realms may be
needed. In section 4, it was argued that quantum entanglement may need yet another such
extension. \\
And as suggested, such possible extensions of the concept of Physical Realms need not
necessarily be given by one and the same additional realm. \\

In case such a multiplicity of realms, beyond the two classical Cartesian ones, may raise
certain concerns, one can always remember that, as thinking humans, thus in particular,
physicists, our basic realm is in fact the "res cogitans". No wonder that Descartes insisted
on what he considered as the fundamental ontological fact for us humans, namely, "cogito,
ergo sum" ... \\
And therefore, without much further intellectual effort, we may at a certain stage subsume all
other possible realms to that one. In other words, we may as well consider that everything is
but a model, including what for so long we considered as having "objective" existence,
whatever "objective" may happen to mean, namely, the Physical Realms. \\

The only major difference such a subsummation may imply is that we should redefine accordingly
what we mean by "experimental evidence", and in particular, by "falsifiability". \\

\end{document}